# Onset of rebound suppression in non–Newtonian droplets post–impact on superhydrophobic surfaces

**Purbarun Dhar\*, 2, Soumya Ranjan Mishra** and **Devranjan Samanta\*, 1**

Department of Mechanical Engineering, Indian Institute of Technology Ropar, Rupnagar–140001, India

\**Corresponding authors*:
[1] E-mail: devranjan.samanta@iitrpr.ac.in
[1]Tel: +91-1881-24-0109
[2]E-mail: purbarun@iitrpr.ac.in

**Abstract**
Droplet deposition after impact on superhydrophobic surfaces has been an important area of study in recent years due to its potential application in reduction of pesticides usage. Minute amounts of long chain polymers added to water has been known to arrest the droplet rebound effect on superhydrophobic surfaces. Previous studies have attributed different reasons like extensional viscosity, dominance of elastic stresses or slowing down of contact line in retraction phase due to stretching of polymer chains. The present study attempts to unravel the existence of critical criteria of polymer concentration and impact velocity on the inhibition of droplet rebound. The impact velocity will indirectly influence the shear rate during the retraction phase, and the polymer concentration dictates the relaxation timescale of the elastic fluids. Finally we show that the Weissenberg number (at onset of retraction), which quantifies both the elastic effects of polymer chains and the hydrodynamics, is the critical parameter in determining the regime of onset of rebound suppression, and that there exists a critical value which determines the onset of bounce arrest. The previous three causes, which are manifestations of elastic effects in non-Newtonian fluids, can be related with the proposed Weissenberg number criterion.

I. Introduction

Research on impact dynamics of non-Newtonian droplets has garnered momentum since the seminal paper by Bergeron et al [1]. Addition of minute amount of flexible polymers in Newtonian base fluid without significant change in viscosity causes suppression of droplet rebound after impact on superhydrophobic (SH) surfaces. Initially it was proposed that the high elongational viscosity of non-Newtonian fluids decelerates the retraction velocity of the droplet after the impact, which in turn suppresses the rebound phase.

However the role of extensional viscosity was ruled out, as it will affect both spreading and retraction phases, whereas experimental observations show that only retraction behaviour is changed due to polymer effects [2, 3]. Subsequently Bonn et al [4] proposed the dominant role of elastic normal stresses over the surface tension, which is the driving force for drop retraction. Another study [5] proposed that the contact line is slowed down during retraction due to the stretching of the flexible polymer chains near the droplet edge. Suppression of droplet rebound on hydrophobic surfaces has potential applications in reduction of pesticides during spraying and reducing environmental damage due to ground contamination of pesticides [1]. Other industrial applications like inkjet printing, slot coating [6], spray coating and cooling [7] may be improved from the studies of non-Newtonian droplet impact dynamics.

Droplet rebound suppression is also achieved by addition of specific surfactants [8], although generation of smaller droplets and splashing is not always inhibited. Application of electric field on non-axisymmetric droplets was also found effective towards inhibiting droplet rebound [9-10]. Some other studies have focussed on spreading and splashing behaviour of droplets of yield stress fluids and shear thickening nature due to modulation in microstructures of hydrophobic surfaces [11-13]. For shear thickening fluid, irrespective of impact velocity, the droplet freezes into a deformed state while spreading, thereby arresting all subsequent dynamics like fragmentation or rebound typical of Newtonian fluid droplets.

Earlier works [1, 4, 5] may inadvertently generate an impression that any amount of polymer or impact velocity will cause the droplet to stick to the hydrophobic surface during retraction. The present paper aims to highlight the role of critical polymer concentration and impact velocity on the onset of droplet rebound suppression for non-Newtonian droplets. We hypothesize that the onset of inhibition of the rebound is dependent on the relaxation time (manifested though the polymer concentration in water) and the shear rate at the contact line during the retraction phase. We will present the potential role of the Weissenberg number in governing the droplet rebound inhibition. The role of extensional viscosity on the longevity of the thin filaments formed during droplet rebound process is also highlighted.

## II.  Experimental setup

Figure 1 shows the experimental setup used in our studies. Experiments were performed with a digitized precision droplet dispenser setup [14]. Impact velocity of the droplets was changed by changing the height of discharge (from a microliter syringe). The superhydrophobic (SH) surfaces were made by spray coating (Ultratech Ever-dry, USA). Aqueous solutions of Polyacrylamide (PAAM ~5 million molecular weight, Sigma Aldrich) of different concentrations were used as working fluids. Homogeneous solutions were prepared using a magnetic stirrer for two hours. Image acquisition was done by a high speed camera (FastCam SA4, Photron) at 3600 fps. Rheological parameters like shear viscosity, viscoelastic moduli and relaxation time were measured by a rotational rheometer (Anton Paar) with parallel plate geometry. Surface tension of the liquids was determined by the pendant drop method. Minimal changes in the surface tension were noted (within ±3-5 % of the surface tension of water). The static contact angles were determined from image analysis and no appreciable change in the wetting state due to addition of polymer is noted (the change in the static contact angle was within 5-7$^o$ compared to the water droplet).

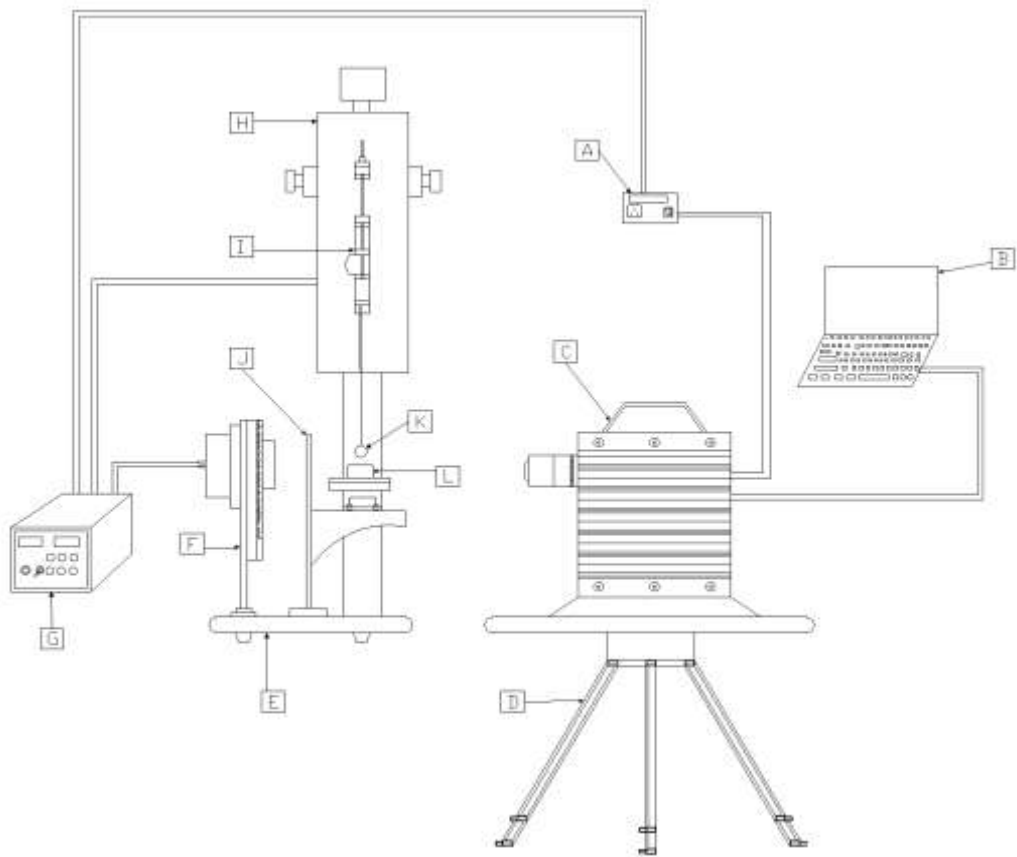

Fig.1 Schematic of the experimental setup (A) Power supply (B) Laptop (C) High speed Camera (D) Tripod (E) Base (F) Back light Arrangement (G) Drop Dispenser Controller (H) Dispenser (I) Syringe Holder (J) Substrate Holder (clamp) (K) Pendant droplet (L) substrate inclination apparatus

### III.     Results and discussions

We start with the case where droplet rebound is suppressed. Figure 2 shows the spatio-temporal evolution of a 1500 ppm PAAM solution droplet on a SH surface. Fig. 2 (a) shows the maximum spreading of the diameter after impact. During retraction (fig. 2 c-f), the droplet gets attached to one location, exhibiting thin filament formation at the ground and mushroom like cloud structure at the top. The thin filament has a short lifetime (~few milliseconds) before the mushroom like structure eventually settles down to the ground due to gravity (fig. 2 g-h). Rheological measurements reveal that the dilute PAAM solutions exhibit augmented elastic moduli with negligible increase in the shear viscosity (typically Boger fluids). The transient thin filament formation is observed due to the viscoelastic nature of the fluids.

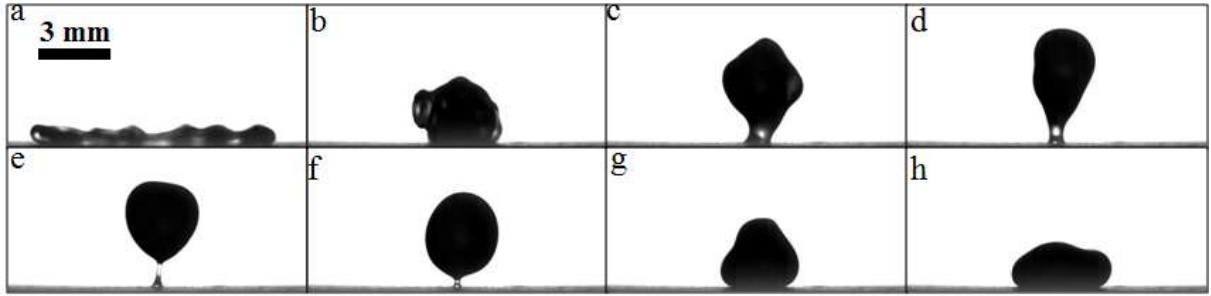

**Fig.2**. Evolution of a droplet of 1500 ppm aqueous PAAM solution post-impact on a SH surface, with impact velocity of 2.25 m/s. The images are temporally spaced 5.56 ms apart.

### A. Role of Polymer concentration

Subsequently we studied the role of polymer concentration on the droplet rebound dynamics at a fixed velocity (refer fig. 3). We have fixed the impact velocity at 2.25 m/s for different polymer concentrations. For low polymer concentrations (50 ppm), droplets rebound like Newtonian cases, albeit with lower degree of fragmentation. Even a very short-lived filament formation was noticed for both 50 ppm ((3-c)) and 200 ppm (3-h & 4 iii). As we increase the polymer concentration, the filament survives long enough to allow the gravitational settling down of the top heavier portion, thereby arresting the droplet rebound off the surface. In our studies, droplet rebound is achieved with a critical polymer concentration of approximately 500 ppm. However, this is only true for impact velocities of 2.25 m/s and above.

In addition, reduced fragmentation into secondary droplets was observed for 50 and 200 ppm (refer fig. 4). Reduced fragmentation was reported earlier in cases of polymeric jets [18-19]. With increasing polymer concentration, thin filaments were formed at the droplet rim during retraction, leading to generation of lower number of secondary smaller droplets. Also, detachment of the secondary droplets was arrested and the filaments recoiled to retract the secondary droplets to fuse with the mother droplet. Top views of drop impact dynamics of water and aqueous solutions of PAAM (100 & 200 ppm) are presented for qualitative comparison in fig. 4. For the same time interval, it can be readily observed that the thin filaments formed during splashing are intact in case of the non-Newtonian fluids, whereas the water droplet exhibits fragmentation into secondary droplets (which tend to scamper away from the impact spot). This accentuates the idea of adding minute amounts of such polymers to reduce pesticide and fertilizer wastage during crop spraying.

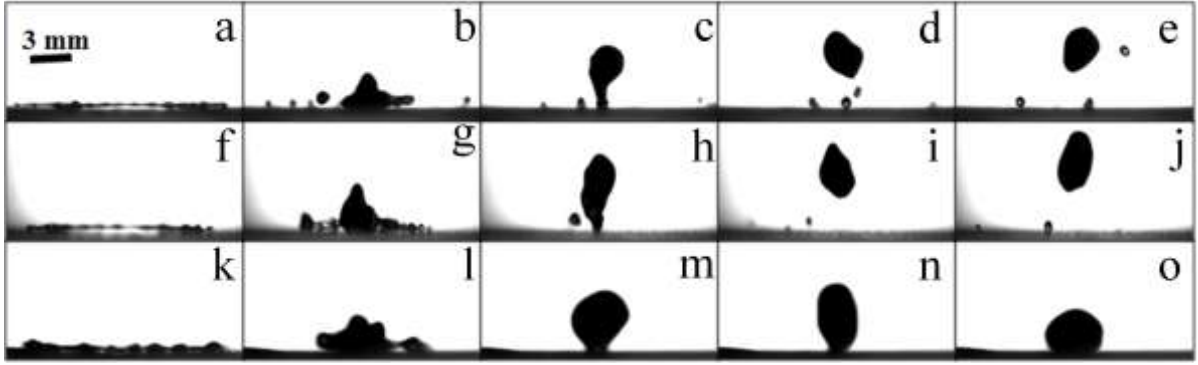

**Fig.3**. Influence of polymer concentration; (a-e): 50 ppm; (f-j): 200 ppm; (k-o): 500 ppm PAAM solution droplet impact dynamics on SH surface with an impact velocity of 2.25 m/s. The images are temporally spaced by a time interval of 6.95 ms. Figure e and j doesn't represent the highest height achieved after rebound.

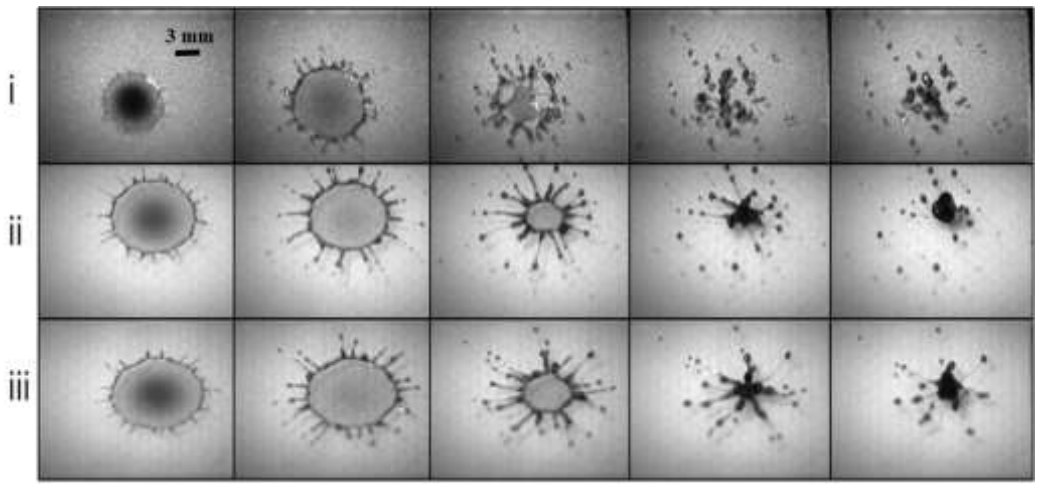

**Fig.4.** Top view images for (i) water (ii) 100 ppm and (iii) 200 ppm PAAM droplets with impact velocity of 2.25 m/s; addition of polymer showing the inhibition of fragmentation of smaller droplets and enhancement of filaments formed during retraction phase. Clear indication of the role of the elastic nature of the fluid can be noted by comparing the pictures in the 3$^{rd}$ column of each row.

### B. Role of shear rate

Then we tried to probe the role of shear rate which also influences the viscoelastic effects and is directly dependent on impact velocity. Viscoelastic effects due to stretching and coiling of polymer chains become more prominent at higher shear rates. Higher shear rates (at the moment of retraction) are achieved with higher velocities at the moment of impact. Even for the highest polymer concentration tested, there is a certain critical velocity below which the shear rates are not strong enough to induce viscoelastic effects for suppressing the droplet rebound. Keeping the polymer concentration constant at 1500 ppm, we varied the impact velocity. Fig. 5 shows the effect of impact velocity (and hence shear rates) on the droplet rebound. At the lowest velocity (fig 5, row (i)), the droplet rebounds without trace of filament formation. With increased velocity (fig 5 (ii), c-f), filament formation is observed both at the ground and between the two parts of the drop. The subsequent shapes resemble bowling pins (fig. 5 (ii)-c), or mushrooms or balloons (fig. 5 (ii) e or f) or wine glasses (fig.5 (iii)-e and (iv)-e). Such droplet shapes are never achieved with Newtonian fluids. With increasing velocity, droplet rebound suppression occurs quicker within the timeframe presented in this figure (fig. 5, row (iv)). Filament dynamics was observed but no bead formation/blistering were apparent, possibly due to lower strain rates unlike pure extensional rheological flows of non-Newtonian fluids [15-18].

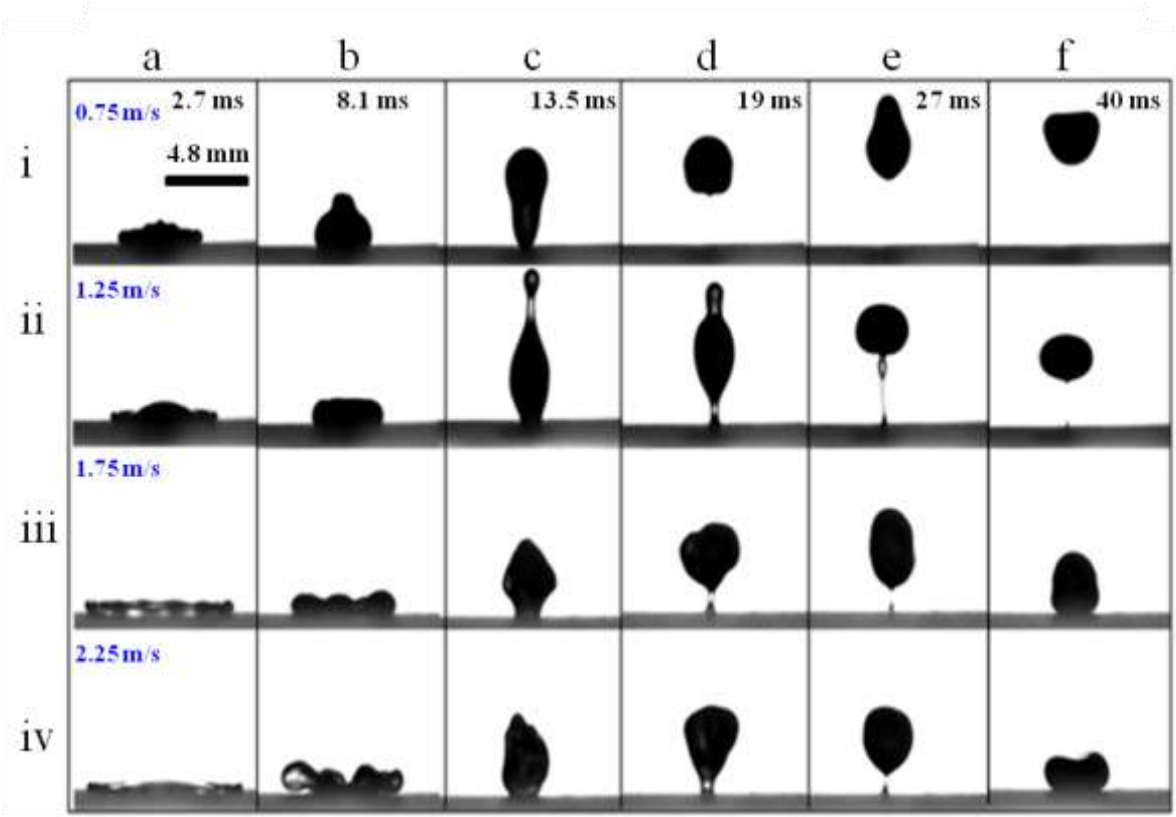

**Fig.5**: Influence of impact velocity on droplet dynamics of 1500 ppm PAAM solution. Left to right shows the temporal evolution and top to bottom shows impact velocity. The propensity of rebound suppression improves with increasing impact velocity. Movies of experiments described in row (i) and (iv) are in movie format in the supplementary material section.

Higher longevity of the thin filaments formed at the base may lead to higher probability of droplet rebound suppression. The elastic recoil of the filaments counters the surface energy component which promotes anti-wetting (subsequent ejection off the surface). In a previous study [15] it was shown that micron scale thick filaments between two beads at higher concentrations (1000 ppm Polyethylene oxide PEO) never break. Likewise, in our case, the filaments connecting the surface and the upper mushroom like portion exhibit higher lifetimes with increasing polymer concentration. Previous studies [19] suggest that increase of polymer concentration causes enhancement of extensional viscosity, leading to enhanced stability of thin filaments formed at the base. The role of extensional viscosity during the formation of filament pinned to the ground can be explored in future studies with capillary (extensional) rheometer facilities. It is worth mentioning that preliminary experiments on inclined planes showed the existence of even longer filaments (with large life times) formed from the base. Although the upper heavier part is attached to the filament for a considerable time in the context of droplet impact dynamics, eventually the larger portion gets detached from the filament and rolls down along the inclined plane due to gravitational potential [refer fig. 6].

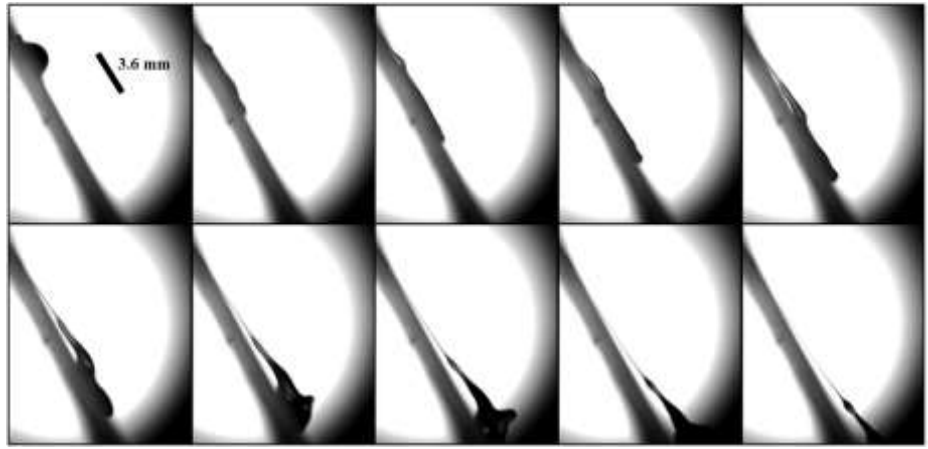

**Fig.6**: Impact dynamics of 1500 ppm on an inclined SH surface at $60^o$ w.r.t. ground. Impact velocity is 2.25 m/s. The array highlights the role and life time of the filaments in case of elastic fluid impact, and its subsequent role (in horizontal case) in suppressing the rebound off the surface. Images were taken at 1.39 ms interval. Movie is provided in the supplementary material section.

### C. Role of Weissenberg number

Based on the figures 2-6, it is evident that both polymer concentration and impact velocity play governing roles to trigger the onset of droplet rebound suppression. Here we present the role of impact velocity vs. polymer concentration (refer fig. 7a), and it is clearly evident that the critical velocity decreases with increasing polymer concentration. Impact velocity is the deciding factor behind the shear stress acting at the contact line during onset of retraction of the droplet. The focus is put on the shear during retraction phase as earlier reports have established that the spreading phase [5] is not altered by the elastic effects.

In order to emphasize the effect of viscoelasticity, we used a brine solution (aqueous NaCl) of viscosity almost similar to the high shear Newtonian viscosity value of the 2000 ppm PAAM solution. At the highest velocity of 2.5 m/s, the droplet of brine undergoes fragmentation into tiny droplets (very similar to water) showing no signature of rebound suppression (refer fig. 8). This signifies that the suppression of fragmentation and rebound is essentially caused by the viscoelastic effect. Fig. 8 shows the usual fragmentation observed in Newtonian solutions like fig. 4(i) without any droplet rebound suppression (for the brine solutions). It is also noteworthy that surface tension of the salt solution and polymer solution are very similar, and the static contact angles are also nearly equal. Hence the viscoelastic effects are responsible for the suppression of the fragmentation and rebound.

Experiments have revealed that for initiation of rebound suppression, there exist parallel criteria of critical impact velocity and critical polymer concentration. We propose that the product of shear rate (localized to the vicinity of the contact line, at the instant of droplet retraction) and the viscoelastic relaxation time of the polymer solutions, better known as the Weissenberg number (Wi), is the driving entity behind this droplet suppression behaviour. Shear rate during retraction was calculated as spreading dynamics was hardly affected due to polymers [1, 4]. Shear rates were estimated by image analysis of the contact line during droplet retraction images using ImageJ software. The typical shear rates at retraction of different impact velocity conditions have been tabulated in table 1. The critical shear rate for rebound suppression decreases with polymer concentration.

Table 1: Relaxation times and shear rate at retraction (for impact at the critical velocity) for different elastic fluid droplets

| Polymer concentration (ppm) | Relaxation time (ms) | Critical impact velocity (m/s) | Shear rate at retraction ($s^{-1}$) |
|---|---|---|---|
| 500 | 3.105 | 2.50 | 431.14 |
| 750 | 4.836 | 2.50 | 356.25 |
| 1000 | 6.567 | 2.25 | 334.90 |
| 1250 | 5.663 | 1.75 | 300.14 |
| 1500 | 4.758 | 1.75 | 337.90 |
| 1750 | 6.896 | 1.25 | 291.69 |
| 2000 | 9.034 | 1.25 | 263.30 |

With the increase of polymer concentration, the relaxation time will increase, and is determined from the frequency dependent viscoelastic moduli and the relaxation modulus using reported standard rheological analysis [20]. The viscoelastic moduli are determined at different shear conditions (using oscillatory frequency modulation at low shear amplitudes of ~ 1 $s^{-1}$). The low shear amplitude is chosen as the dilute PAAM solutions used show non-Newtonian behaviour only up to ~1 $s^{-1}$ in the steady shear rheology. The complex viscosity is determined from the frequency dependent viscoelastic moduli and effective relaxation times are deduced from standard methodology reported in literature [20]. The typical relaxation time scales for the used PAAM solutions are tabulated in table 2.

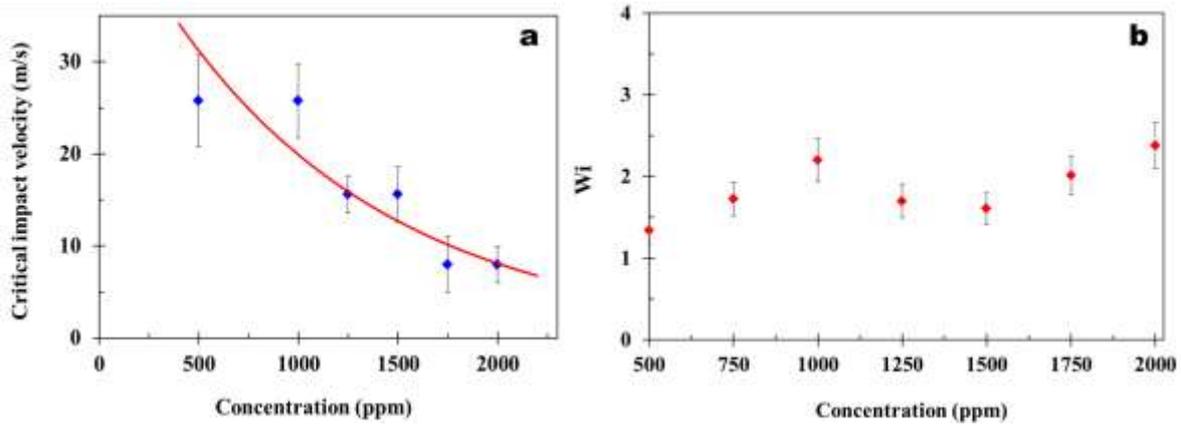

**Fig.7:** (a) Critical impact velocity vs. critical polymer concentration (ppm) (b) Critical Weissenberg number (Wi) vs. polymer concentration (ppm).

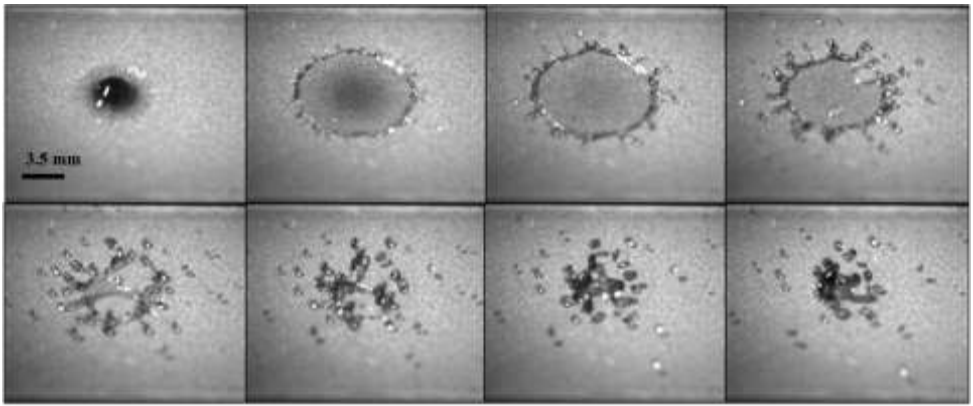

**Fig.8**: Impact dynamics of salt water (viscosity ~ 1.5 times of water, similar to Newtonian viscosity of 2000 ppm PAAM solution). Images were taken at 1.39 ms interval.

For the relaxation time scales, agreement with reports on viscoelastic behaviour of PAAM solutions [21] was noted. Consequently, the impact velocity for which the rebound is arrested also decreases with the polymer concentration. This signifies that unlike proposed earlier, it will be possible to arrest rebound of droplets with vanishing amounts of polymer, if the impact velocity is sufficiently large (assuming that the infinitely dilute polymer solution can sustain the fragmentation due to high inertia impact, and the polymer dissolved has sufficiently long chains to ensure Wi >1 even at infinite dilution) . Fig. 7 (b) shows the critical Wi vs. polymer concentration. In spite of the uncertainties associated with the experimental materials and methodology, the Wi values can be observed to be confined within a narrow regime of ~ 1.25–2.5. This scenario is identical to earlier studies where it was shown that for Wi>1 elastic instabilities sets in [22]. Although the present scenario does not lead to elastic turbulence [23] or elasto-inertial turbulence [24], suppression of non-Newtonian droplets rebound is clearly a viscoelastic phenomenon dependent on critical value of Wi. High precision experiments or numerical simulations may be used to identify a single unique value of Wi above which droplet rebound suppression will occur on SH surfaces.

However, as all the values lie in Wi > 1, it is confirmed that the elastic nature of the fluid plays a vital role towards arrest of the rebound behaviour.

## IV. Conclusions

To conclude, in this work, we show the existence of parallel criteria of critical shear rate at retraction and polymer relaxation time on the onset of droplet rebound inhibition on SH surfaces. Previous studies focused on extensional viscosity [1], normal stresses [4] or stretching of polymer molecules at the receding contact line [5]. All these reasons are basically different manifestations of viscoelasticity, which is absent for Newtonian fluids. Our studies have shown that viscoelastic effects (majorly the relaxation timescale) are modified due to variation of polymer concentration and the shear rates during the retraction phase (which can be directly modulated by the impact height). Finally we showed that onset of droplet rebound suppression is observed for a critical threshold Weissenberg number. The critical Wi is noted to be bound within a narrow regime for the spectrum of impact velocities and polymer concentrations studied. We believe that with experiments of higher precision (or intricate computations) a unique Wi value for properly characterized SH surfaces can be obtained, which triggers the onset of rebound inhibition. In addition, the temporal evolution of drops during retraction and rebound will be challenging phenomena for numerical simulations. Further, force measurement during impacts of non-Newtonian droplets can lead to greater insight [25]. Our studies maybe beneficial in deciding the optimum amount of non-Newtonian additives and release height for maximizing the droplet depositions on hydrophobic substrates.

## Acknowledgements
SRM would like to thank the Ministry of Human Resource Development, Govt. of India, for the scholarship. DS and PD would like to thank IIT Ropar for funding the present work (vide grants 9-246/2016/IITRPR/144 & IITRPR/Research/193 respectively).